\documentclass[%
 aip,
 amsmath,amssymb,
 reprint,%
]{revtex4-1}

\usepackage{graphicx}
\usepackage{dcolumn}
\usepackage{bm}
\usepackage{euscript,stmaryrd}
\usepackage{xcolor}
\usepackage[utf8]{inputenc}
\usepackage[T1]{fontenc}
\usepackage{mathptmx}

\begin{document}

\preprint{AIP/123-QED}

\title[Stable chimeras of non-locally coupled Kuramoto--Sakaguchi oscillators in a finite array]
{Stable chimeras of non-locally coupled Kuramoto--Sakaguchi oscillators in a finite array}

\author{Seungjae Lee}
 \affiliation{Department of Physics, Chonbuk National University, Jeonju 54896, Korea}
\author{Young Sul Cho}%
 \email{yscho@jbnu.ac.kr}
\affiliation{Department of Physics, Chonbuk National University, Jeonju 54896, Korea}
\affiliation{Research Institute of Physics and Chemistry, Chonbuk National University, Jeonju 54896, Korea}

\date{\today}

\begin{abstract}
We consider chimera states of coupled identical phase oscillators
where some oscillators are phase synchronized while others are desynchronized.
It is known that chimera states of non-locally coupled Kuramoto--Sakaguchi oscillators in arrays of finite size 
are chaotic transients when the phase lag parameter $\alpha \in (0, \pi/2)$; 
after a transient time, all the oscillators are phase synchronized, 
with the transient time increasing exponentially with the number of oscillators.
In this work, we consider a small array of six non-locally coupled oscillators 
with the phase lag parameter $\alpha \in (\pi/2, \pi)$ in which the complete phase synchronization of the oscillators
is unstable. Under these circumstances, we observe a chimera state spontaneously formed by the partition of oscillators
into two independently synchronizable clusters of both stable and unstable synchronous states.
We provide numerical evidence supporting that the
instantaneous frequencies of the oscillators of the chimera state are periodic functions of time with a common period,
and as a result, the chimera state is stable but not long-lived transient. We also measure the basin stability of the chimera state and
show that it can be observed for random initial conditions when $\alpha$ is slightly larger than $\pi/2$.
\end{abstract}

\maketitle

\begin{quotation}
A chimera state is the partition of coupled indistinguishable oscillators 
into two subsets with distinct behaviors (coherent and incoherent).  
It has been shown that a stable chimera of non-locally coupled Kuramoto--Sakaguchi oscillators in arrays
with the phase lag parameter $\alpha \in (0, \pi/2)$ exists in the thermodynamic limit. However, the chimera state becomes unstable 
as the number of oscillators becomes finite, and as a result, it collapses to complete phase synchronization after a certain transient time.
In this paper, we numerically show that a stable finite-sized chimera state exists
if complete phase synchronization is avoided by taking $\alpha \in (\pi/2, \pi)$. 
\end{quotation}

\section{\label{sec:introduction} Introduction}

The chimera state, a phenomenon where coupled identical oscillators are partitioned into 
coherent and incoherent subsets~\cite{abrams_chimera_review, omelchenko_chimera_review},
has been widely studied both theoretically~\cite{kuramoto_chimera, abrams_prl_2004, abrams_ring_2006, abrams_prl_2008,  chimera_chaotic, chimera_transient, chaos_weak_chimera1, chaos_weak_chimera2, weak_chimera_chaotic, weak_chimera_chaotic2, chimera_persistent1, chimera_persistent3, chimera_persistent4, yscho_prl_2017, chimera_spectral, pikovsky_avoidcomplete,  nonlocal_infinite, finite_Sakaguchi, loss_of_coherence} and experimentally~\cite{Hagerstrom2012, Tinsley2012, PNAS_chimera, star_chimera, Hart2016, chimera_transient_exp, exotic_nanoelectromechanical, chimera_persistent2, transient_chaotic_exp, coupled_pendula, not_multiplicity_laser, laser_chimera} using various definitions of coherence and incoherence~\cite{chimera_classification}.
The first observation of a chimera state was in arrays of non-locally coupled Ginzburg--Landau oscillators~\cite{kuramoto_chimera}.
In the state, oscillators in an array are partitioned into two domains: one composed of 
phase-locked (coherent) oscillators, and one composed of drifting (incoherent) oscillators.
To understand the phenomenon analytically, non-locally coupled Kuramoto-Sakaguchi oscillators~\cite{kuramoto_sakaguchi} in arrays with
the phase lag parameter $\alpha \in (0, \pi/2)$ have been employed, with which it has been shown that a stable chimera state exists
in the limit of an infinite number of oscillators $N \rightarrow \infty$~\cite{abrams_prl_2004, abrams_ring_2006, nonlocal_infinite}. 
However, it was later reported that the chimera state becomes chaotic transient with finite $N$~\cite{chimera_chaotic, chimera_transient, chimera_spectral}, because
the complete phase synchronization of all the oscillators is stable in the range $0 < \alpha < \pi/2$ such that the chimera state
collapses to the complete phase synchronization after a transient time.
Here, the transient time increases exponentially with $N$~\cite{chimera_transient, chimera_transient_exp, finite_Sakaguchi}, which is consistent with the analytical result that 
the chimera state is stable in the limit $N \rightarrow \infty$~\cite{abrams_prl_2004, abrams_ring_2006, nonlocal_infinite}.

In this paper, we consider an array of six non-locally coupled Kuramoto--Sakaguchi oscillators with the phase lag parameter $\alpha \in (\pi/2, \pi)$, where complete phase synchronization is unstable and thus avoided.
With this setup, we numerically observe a chimera state in which two oscillators are phase synchronized (coherent)
while the other four oscillators are desynchronized (incoherent). Here, phase synchronization of the two oscillators is guaranteed
because they receive the same input from the other four oscillators by permutation symmetry~\cite{remote_prl_2013, pecora_ncomm_2014, pecora_sciadv_2016, yscho_prl_2017}.
Moreover, we show numerically that all oscillators behave periodically with a common period, and as a result, 
the four incoherent oscillators maintain their desynchronization, thereby leading to the chimera state being stable but not long-lived transient. 
We note that chimera states with $\alpha \in (0, \pi/2)$ would collapse rapidly for such a small number of oscillators ($N=6$)~\cite{chimera_transient, chimera_transient_exp, finite_Sakaguchi}.

There have been several approaches to find stable chimeras with finite $N$ by 
changing oscillators and coupling structures~\cite{ chaos_weak_chimera1, chaos_weak_chimera2, weak_chimera_chaotic, chimera_persistent1, chimera_persistent2, chimera_persistent3, chimera_persistent4, yscho_prl_2017}. 
Our approach in this paper claims that the avoidance of complete phase synchronization is key to observe a stable chimera state composed of 
a finite number of oscillators~\cite{yscho_prl_2017, pikovsky_avoidcomplete, not_multiplicity_laser, loss_of_coherence}.

The rest of this paper is organized as follows. In Sec.~\ref{sec:observe_chimera}, we describe a dynamical system where we observe a chimera state
and identify the underlying mechanism of its formation. In Sec.~\ref{sec:persistence_chimera},
we present numerical evidence to support that the instantaneous frequencies of the oscillators of the observed chimera state are periodic functions of time with a common period.
In Sec.~\ref{sec:basin_chimera}, we measure the basin stability~\cite{basin_stability} of the chimera state and other possible states in the system, and in Sec.~\ref{sec:discussion}, we discuss the chimera state from the perspective of frequency synchronization and show that
it is a weak chimera state~\cite{chaos_weak_chimera1, chaos_weak_chimera2, weak_chimera_chaotic, weak_chimera_chaotic2}.

\section{\label{sec:observe_chimera} Observation of a chimera state}
\subsection{Kuramoto--Sakaguchi oscillators in a given network}
We consider the Kuramoto--Sakaguchi model of phase oscillators~\cite{kuramoto_sakaguchi}. 
In this model, the time derivative of the phase of each oscillator in a network is given by
\begin{equation}
\dot{\phi_i}(t)= \omega_i + K\sum_{j=1}^N A_{ij} \textrm{sin}(\phi_j(t)-\phi_i(t)+\alpha)
\label{Eq:governing_eq_origin}
\end{equation} 
for global coupling strength $K > 0$ and phase lag parameter $\alpha \in (0, \pi)$, where $\phi_i \in [0, 2\pi)$ ($i=1,...,N$) is the phase of the $i$-th oscillator and $A_{ij}$ is each entry of $N \times N$ adjacency matrix $\bold A$ of the network. 

We let all oscillators be {\it {identical}} such that they have the same natural frequency $\omega_i = \omega$ for $\forall i$.
If we use a rotating reference frame $\phi_i \rightarrow \phi_i + \omega t$ for $\forall i$ and time scaling $t \rightarrow t/K$,
Eq.~(\ref{Eq:governing_eq_origin}) has the form
\begin{equation}
\dot{\phi_i}(t)= \sum_{j=1}^N A_{ij} \textrm{sin}(\phi_j(t)-\phi_i(t)+\alpha).
\label{Eq:governing_eq_final}
\end{equation}

To observe a chimera state in a finite array of non-locally coupled identical oscillators, 
we use a network of $N=6$, as depicted in Fig.~\ref{Fig:chimera_observation}(a), 
where each oscillator is coupled with neighbors within distance two on the ring. 
In this paper, we use Eq.~(\ref{Eq:governing_eq_final}) with $A_{ij}$ of the network to find the chimera state. 
 
\subsection{\label{cs_pattern} Partition of network oscillators into two independently synchronizable clusters} 

\begin{figure}[t!]
\includegraphics[width=1.0\linewidth]{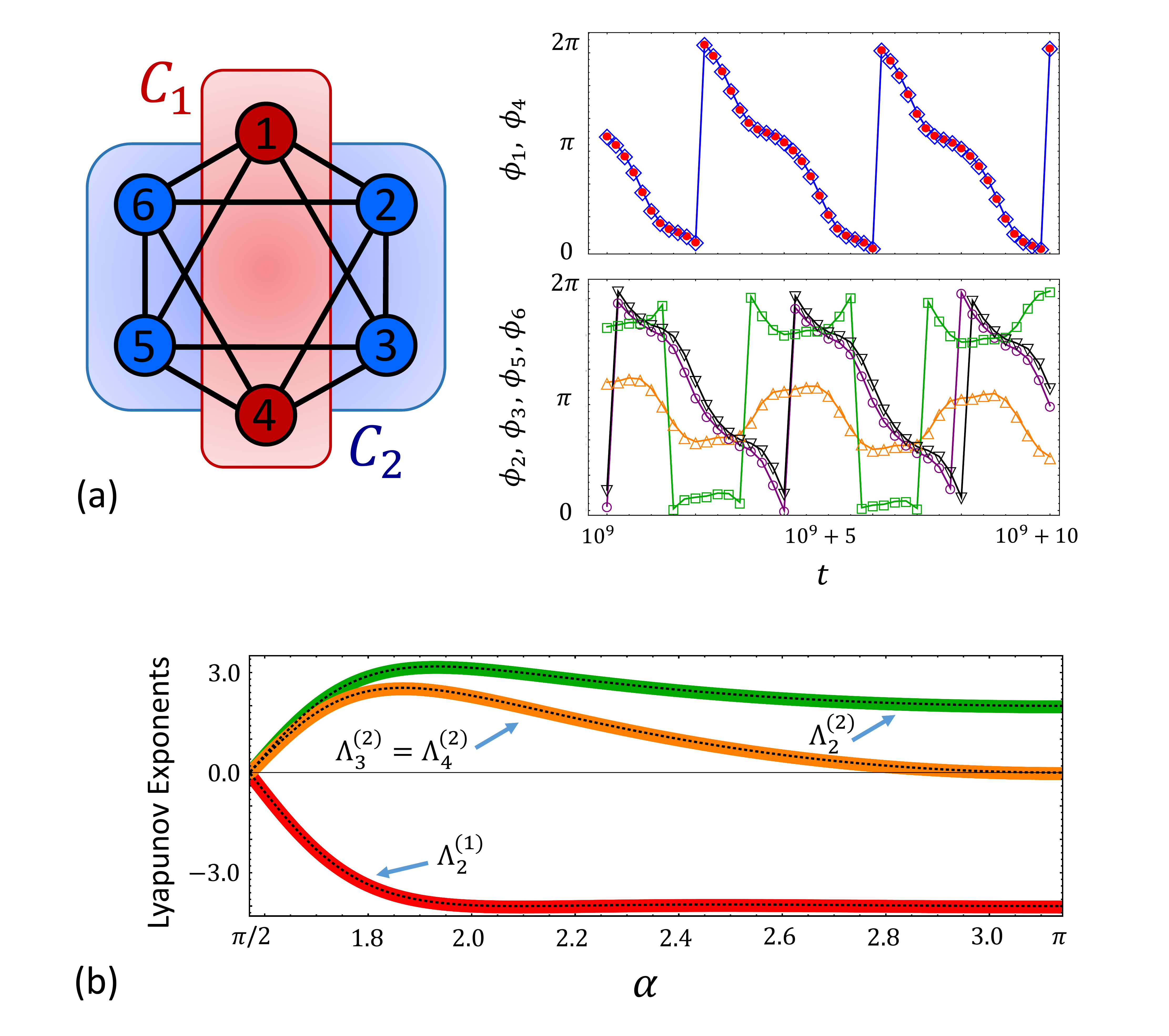}
\caption{(a) Left: Schematic diagram of the network used in this paper. 
Right: $\phi_1(\diamond)$, $\phi_4(\bullet)$, $\phi_2(\circ)$, $\phi_3(\oblong)$, $\phi_5(\triangledown)$, and $\phi_6(\vartriangle)$ of the chimera state observed in the network.
To observe this state, we integrate Eq.~(\ref{Eq:governing_eq_final}) with $\alpha=1.58$ 
for a random initial condition $(\phi_1,...,\phi_6) \in [0, 2\pi)^6$ at $t = -10^3$
to set $t$ to zero after an initial transient. (Note that the chimera state is observed for $t \geq 0$ in 
Figs.~\ref{Fig:chimera_periodicity} and \ref{Fig:chimera_stability}.)
(b) Thick lines indicate numerically estimated transverse Lyapunov exponents $\Lambda^{(m)}_{\kappa}$ for each cluster $C_m$.
To obtain these lines, we integrate Eq.~(\ref{Eq:variational_eq_C1C2}) with Eq.~(\ref{Eq:quotient_C1C2}) up to $t=10^5$ for each $\alpha$.
To discard the initial transient, we numerically integrate Eq.~(\ref{Eq:quotient_C1C2}) over $-10^5 \leq t \leq 0$ 
for randomly taken $s_m(-10^5) \in [0, 2\pi)$ $(m=1,2)$ to obtain $s_1(0)$ and $s_2(0)$ for each $\alpha$.
Dotted lines indicate the functional form of $\Lambda^{(m)}_{\kappa}(\alpha)$ discussed in the main text.}
\label{Fig:chimera_observation}
\end{figure}

The six oscillators in Fig.~\ref{Fig:chimera_observation}(a) are partitioned into two clusters $C_1=\{1,4\}$ and $C_2=\{2,3,5,6\}$.
We denote the synchronous phase of the first cluster by $s_1$ and that of the second cluster by 
$s_2$. Then, the time derivatives of $s_1$ and $s_2$ are respectively given by
\begin{equation}
\begin{split}
\dot{s}_1(t) &= \textrm{sin}(\phi_2(t)-s_1(t)+\alpha)+ \textrm{sin}(\phi_3(t)-s_1(t)+\alpha)\\ 
&+ \textrm{sin}(\phi_5(t)-s_1(t)+\alpha)+ \textrm{sin}(\phi_6(t)-s_1(t)+\alpha),\\ 
\dot{s}_2(t) &= 2\textrm{sin}(\alpha)+\textrm{sin}(\phi_1(t)-s_2(t)+\alpha)\\&+\textrm{sin}(\phi_4(t)-s_2(t)+\alpha).
\end{split}
\label{Eq:governing_eq_C1C2}
\end{equation}
Therefore, the synchronous phase of each cluster evolves following Eq.~(\ref{Eq:governing_eq_C1C2}), meaning
that each cluster can be synchronous irrespective of oscillator phases of the other cluster.

\subsection{Observation of a chimera state where only one cluster is synchronized }

A chimera state of synchronized $C_1$ and desynchronized $C_2$ is discovered using the following procedure.
(i) We avoid the complete phase synchronization of all six oscillators by using $\alpha \in (\pi/2, \pi)$ in which complete phase synchronization is unstable.
(ii) In this range of $\alpha$, we integrate the quotient network dynamics of Eq.~(\ref{Eq:governing_eq_final}) 
for the two synchronous clusters $C_1$, $C_2$ and show that synchronous state of $C_1$ is stable whereas that of $C_2$ is unstable
along the trajectory of the two synchronous clusters.
(iii) Finally, we observe the chimera state in the range of $\alpha$ by integrating the governing equation (Eq.~(\ref{Eq:governing_eq_final})) numerically
for random initial phases.

We consider the quotient network dynamics of Eq.~(\ref{Eq:governing_eq_final})
for the two synchronous clusters $C_1$, $C_2$ given by
\begin{equation}
\begin{split}
\dot{s}_1(t) &= 4\textrm{sin}(s_2(t)-s_1(t)+\alpha)\\ 
\dot{s}_2(t) &= 2\textrm{sin}(s_1(t)-s_2(t)+\alpha)+2\textrm{sin}(\alpha), 
\end{split}
\label{Eq:quotient_C1C2}
\end{equation}
where $s_1$, $s_2$ are the phases of synchronous clusters $C_1$, $C_2$, respectively (i.e. $s_1=\phi_1=\phi_4$ and $s_2=\phi_2=\phi_3=\phi_5=\phi_6$). 
A variational equation of Eq.~(\ref{Eq:quotient_C1C2}) along the trajectory of complete phase synchronization $s(t)=s_1(t)=s_2(t)$ is given by
\begin{equation}
\dot{\eta}(t)=-6\textrm{cos}(\alpha)\eta(t)
\end{equation}
for $s_1(t)=s(t)-2\eta(t)$ and $s_2(t)=s(t)+\eta(t)$.
We find that $\eta(t)$ diverges for $\pi/2 < \alpha < \pi$ such that 
complete phase synchronization is unstable and therefore avoided. Accordingly, the phases of the two synchronous clusters remain distinct ($s_1(t) \neq s_2(t)$) in the range $\pi/2 < \alpha < \pi$.

\begin{figure}[t!]
\includegraphics[width=1.0\linewidth]{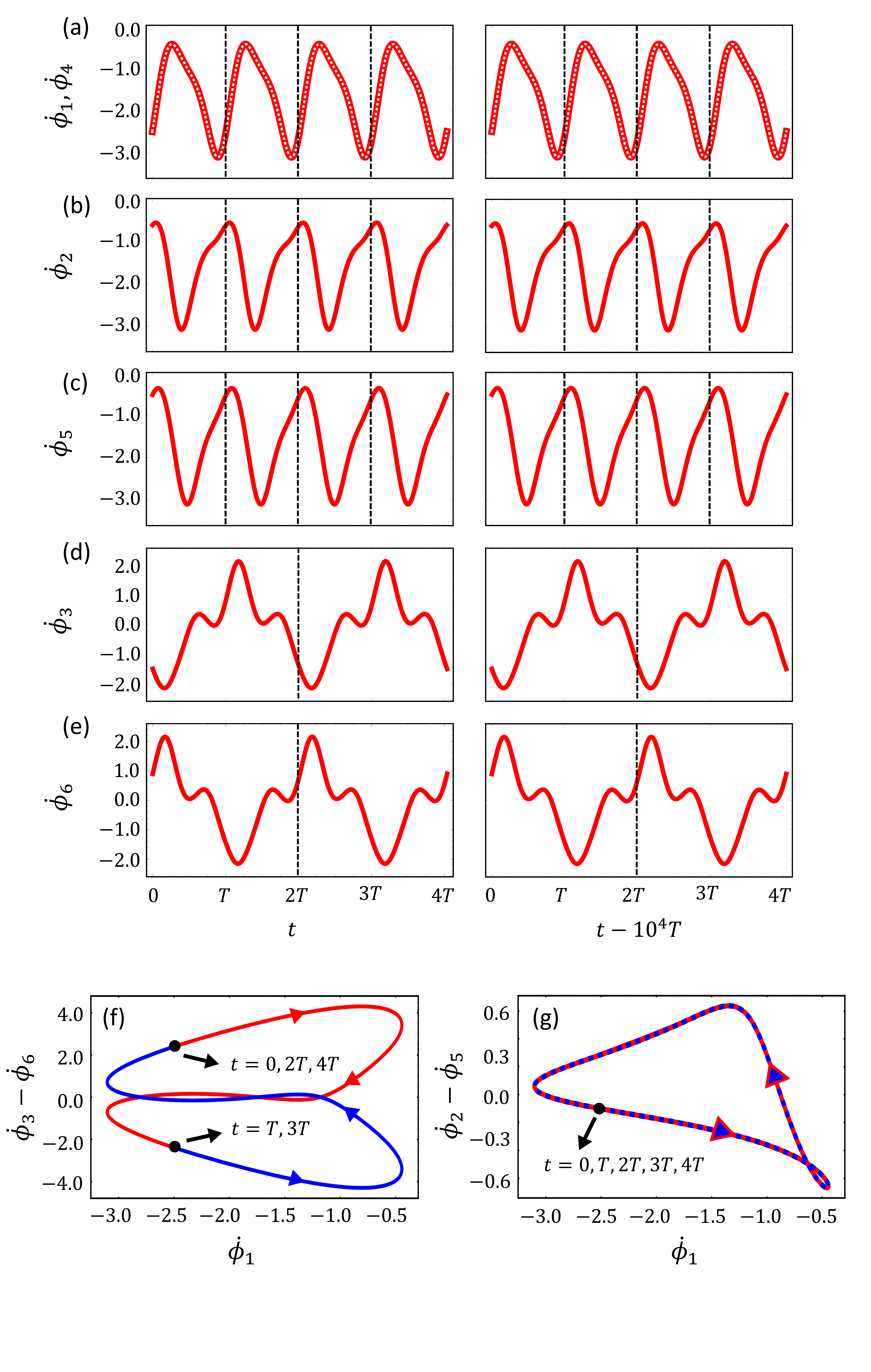}
\caption{Periodicity in the time series of $\dot{\phi}_i(t)$ of the chimera state in Fig.~\ref{Fig:chimera_observation}(a).  
(a--c) Numerical data to support that 
(a) $\dot{\phi}_1$ (solid line) and $\dot{\phi}_4$ (dotted line), 
(b) $\dot{\phi}_2$, and (c) $\dot{\phi}_5$ are periodic functions with period $T$.
Comparison between the left and right panels in each row shows that
the same pattern of $\dot{\phi}_i(t)$ $(i=1,2,4,5)$ during period $T$ appears after $10^4$ cycles.  
(d,e) Numerical data to support that (d) $\dot{\phi}_3$ and (e) $\dot{\phi}_6$ are periodic functions with period $2T$. 
Comparison between the left and right panels in each row shows that
the same pattern of $\dot{\phi}_i(t)$ $(i=3,6)$ during period $2T$ appears after $5 \times 10^3$ cycles.  
(f,g) (f) $(\dot{\phi}_1, \dot{\phi}_3-\dot{\phi}_6)$ and (g) 
$(\dot{\phi}_1, \dot{\phi}_2-\dot{\phi}_5)$ for $\dot{\phi}_i$ in the left panels of (a--e).
$(\dot{\phi}_1, \dot{\phi}_3-\dot{\phi}_6)$ moves around a fixed path two times with period $2T$,
whereas $(\dot{\phi}_1, \dot{\phi}_2-\dot{\phi}_5)$ moves around a fixed path four times
with period $T$. Arrows indicate the direction of motion. 
These results support that the least common multiple of the periods of all $\dot{\phi}_i$ is indeed $2T$.}
\label{Fig:chimera_periodicity}
\end{figure}

\begin{figure}[t!]
\includegraphics[width=1.0\linewidth]{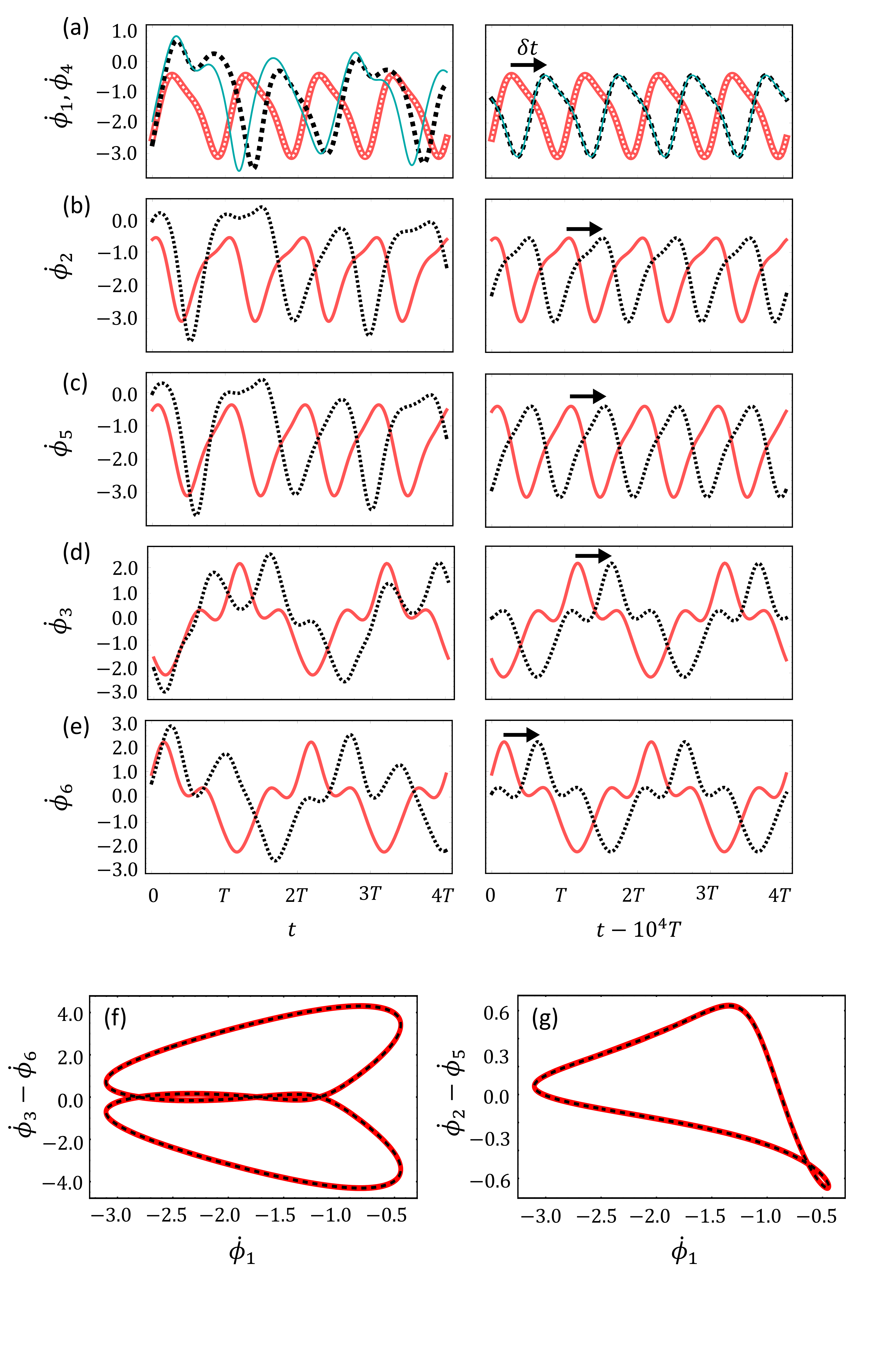}
\caption{Constant time shift $\delta t$ of all $\dot\phi_i(t)$ of the chimera state in Fig.~\ref{Fig:chimera_observation}(a) against initial phase perturbation.
(a) The same $\dot{\phi}_1$, $\dot{\phi}_4$ of the chimera state in Fig.~\ref{Fig:chimera_periodicity}(a),
and $\dot{\phi}_1$ (thick dotted line), $\dot{\phi}_4$ (thin solid line)
of the trajectory perturbed at $t=0$.
(b--e) The same $\dot{\phi}_i$ $(i=2,3,5,6)$ of the chimera state in Fig.~\ref{Fig:chimera_periodicity}(b--e), 
and $\dot{\phi}_i$ (dotted line) of the trajectory perturbed at $t=0$.
(a--e) The initial phase perturbation of each oscillator is given by random numbers $\delta\phi_i(0) \in [-1, 1]$.
In the right panels, each $\dot{\phi}_i$ of the perturbed trajectory is shifted forward
by $\delta t \approx 0.89$ constantly for $\forall i$ compared to those of the chimera state.
(f,g) (f) $(\dot{\phi}_1, \dot{\phi}_3-\dot{\phi}_6)$ and (g) $(\dot{\phi}_1, \dot{\phi}_2-\dot{\phi}_5)$
of the chimera state (solid line) and the perturbed trajectory (dotted line) for $\dot{\phi}_i$ in the right panels of (a--e).
On each plane, 
both trajectories move around the same path,
which supports a constant time shift $\delta t$ of all $\dot{\phi}_i$.}
\label{Fig:chimera_stability}
\end{figure}

Along the trajectory $(s_1(t), s_2(t))$ of Eq.~(\ref{Eq:quotient_C1C2}) for $\pi/2 < \alpha < \pi$,
we show that the synchronous state of $C_1$ is stable whereas that of $C_2$ is unstable. For the deviation of each phase
$\delta\phi_i = \phi_i - s_m$ for $i \in C_m$ $(m=1,2)$, we consider perturbation transverse to the 
synchronization manifold of each cluster. Specifically, we consider perturbations
$\eta^{(1)}_{\kappa}$ $(\kappa=2)$ for $C_1$ and $\eta^{(2)}_{\kappa}$ $(\kappa=2,3,4)$ for $C_2$, 
where $\eta^{(1)}_{2} = (\delta \phi_1 - \delta \phi_4)/\sqrt{2}$, 
$\eta^{(2)}_2 = (-\delta\phi_2+\delta\phi_3-\delta\phi_5+\delta\phi_6)/2$, $\eta^{(2)}_3 = (\delta\phi_2-\delta\phi_5)/\sqrt{2}$,
and $\eta^{(2)}_4 = (\delta\phi_3-\delta\phi_6)/\sqrt{2}$. Then, variational equations of Eq.~(\ref{Eq:governing_eq_final})
along $\phi_i = s_m$ for $i \in C_m$ $(m=1,2)$ are given by
\begin{equation}
\begin{split}
\dot{\eta}_{2}^{(1)}(t) &= -4\textrm{cos}(s_2(t)-s_1(t)+\alpha)\eta_{2}^{(1)}(t), \\ 
\dot{\eta}_{2}^{(2)}(t) &= -2\big[\textrm{cos}(s_1(t)-s_2(t)+\alpha)+2\textrm{cos}(\alpha)\big]\eta_{2}^{(2)}(t), \\ 
\dot{\eta}_{3}^{(2)}(t) &= -2\big[\textrm{cos}(s_1(t)-s_2(t)+\alpha)+\textrm{cos}(\alpha)\big]\eta_{3}^{(2)}(t), \\ 
\dot{\eta}_{4}^{(2)}(t) &= -2\big[\textrm{cos}(s_1(t)-s_2(t)+\alpha)+\textrm{cos}(\alpha)\big]\eta_{4}^{(2)}(t).
\end{split}
\label{Eq:variational_eq_C1C2}
\end{equation}
We numerically obtain transverse Lyapunov exponents 
$\Lambda^{(m)}_{\kappa}=(1/t){\textrm {ln}(||\eta^{(m)}_{\kappa}(t)||/||\eta^{(m)}_{\kappa}(0)}||)$ for $t \gg 1$,
as shown in Fig.~\ref{Fig:chimera_observation}(b). 
We note that $\Lambda^{(2)}_3=\Lambda^{(2)}_4$ from Eq.~(\ref{Eq:variational_eq_C1C2}).

We can obtain a functional form of $\Lambda^{(m)}_{\kappa}$ depending on $\alpha$ 
by using the evolving $s_1(t)$, $s_2(t)$ of Eq.~(\ref{Eq:quotient_C1C2})
with time-independent phase difference $Y=s_1(t)-s_2(t)$.
If we insert $Y=s_1(t)-s_2(t)$ into Eq.~(\ref{Eq:quotient_C1C2}) with $\dot{s}_1=\dot{s}_2$,
we can derive $Y$ as a function of $\alpha$ such that 
$Y(\alpha)={\textrm {cos}}^{-1}\big[(-4-5{\textrm {cos}}(2\alpha))/(5+4{\textrm {cos}}(2\alpha))\big]$.
Then, we obtain a functional form of $\Lambda^{(m)}_{\kappa}(\alpha)$ from Eq.~(\ref{Eq:variational_eq_C1C2}) 
using the relation $\dot{\eta}^{(m)}_{\kappa}(t)=\Lambda^{(m)}_{\kappa}(\alpha)\eta^{(m)}_{\kappa}(t)$.
We check that this analytic form of $\Lambda^{(m)}_{\kappa}(\alpha)$
agrees well with the numerical result, as shown in Fig.~\ref{Fig:chimera_observation}(b).

For $\pi/2 < \alpha < \pi$ where complete phase synchronization (i.e. $\phi_i(t)=s(t)$ for $\forall i$) is avoided, 
we find that $\Lambda^{(1)}_2 < 0$, $\Lambda^{(2)}_2>0$, and $\Lambda^{(2)}_3=\Lambda^{(2)}_4 > 0$ as in Fig.~\ref{Fig:chimera_observation}(b)
such that the synchronous state of $C_1$ is stable while that of $C_2$ is unstable along the trajectory $s_1(t) \neq s_2(t)$ of Eq.~(\ref{Eq:quotient_C1C2}). 
Therefore, we expect that the chimera state can be observed in the range $\pi/2 < \alpha < \pi$ 
using random initial conditions of $\phi_i$ for which only the oscillators in $C_1$ would be synchronized spontaneously.
Via numerical integration of Eq.~(\ref{Eq:governing_eq_final}), we indeed observe that the chimera state persists even after 
$t = 10^{9}$ for a random initial condition with $\alpha \in (\pi/2, \pi)$,
as shown in the right panel of Fig.~\ref{Fig:chimera_observation}(a).  

\section{\label{sec:persistence_chimera} Numerical evidence for $\dot{\phi}_i(t)$ as periodic functions with a common period}

To show that the chimera state in the right panel of Fig.~\ref{Fig:chimera_observation}(a) is stable 
but not long-lived transient, 
we present numerical evidence to support the
periodic behavior of the state. 
Specifically, we obtain numerically that $\dot{\phi}_i(t+T)=\dot{\phi}_i(t)$ $(i=1,2,4,5)$ and
$\dot{\phi}_i(t+2T)=\dot{\phi}_i(t)$ $(i=3,6)$ for $t \geq 0$ with constant $T \approx 2.02$,
as shown in Fig.~\ref{Fig:chimera_periodicity}.
We note that the least common multiple of the periods of all $\dot{\phi}_i$ is $2T$. 
This periodic behavior might be understood analytically 
by finding the integral of motion for this state~\cite{chaos_weak_chimera2}.

As previously mentioned, in the chimera state, two oscillators $C_1=\{1,4\}$ are phase synchronized while the other
four oscillators $C_2=\{2,3,5,6\}$ are desynchronized. 
A necessary condition for the phase synchronization of two oscillators 
$\{i, j\}$ over a (finite) interval of $t$
is $\dot\phi_i(t)=\dot\phi_j(t)$ over the interval of $t$.
Based on the numerical results in Fig.~\ref{Fig:chimera_periodicity},
no pair of oscillators $\{i, j\}$ for $1 \leq i \neq j \leq 6$ satisfies $\dot\phi_i(t) = \dot\phi_j(t)$ over the common period $2T$
except the pair $C_1 = \{1, 4\}$, which repeats every $2T$.
Therefore, the chimera state where only the pair $C_1=\{1, 4\}$ is phase synchronized would persist permanently.

To investigate the linear stability of the trajectory $(\phi_1(t),...,\phi_6(t))$ of the chimera state,
we numerically integrate Eq.~(\ref{Eq:governing_eq_final}) to obtain a perturbed trajectory for $t \geq 0$ 
for a given random initial perturbation of phases $\phi_i(0) \rightarrow \phi_i(0) + \delta\phi_i(0)$,
and then compare the two trajectories for $t \geq 0$.
For the random initial perturbation of phases $\delta\phi_i(0)$ used in Fig.~\ref{Fig:chimera_stability}, 
there is a finite time shift $\delta t$ between time series
$\dot{\phi}_i(t)$ of the two trajectories after an initial transient. 
Therefore, the difference between $\phi_i(t)$ of the two trajectories should be finite as $t \rightarrow \infty$ 
such that the trajectory of the chimera state is unstable or neutrally stable. We checked numerically that the largest nontrivial Lyapunov exponent 
along the trajectory of the chimera state has a small
positive value close to zero, $0.00006 \pm 0.00002$, which supports that the trajectory of the chimera state would be neutrally stable~\cite{lyapunov_chimera}.

However, as shown in Fig.~\ref{Fig:chimera_stability}, we find that time shift $\delta t$ is the same regardless of $i$,
which means that every $\dot\phi_i$ of the perturbed trajectory behaves the same 
as that of the trajectory of the chimera state for time shift $t \rightarrow t - \delta t$.
We also observe a constant time shift of all $\dot\phi_i$ 
with varying $\delta t$ depending on the initial perturbation of phases $\delta\phi_i(0)$.
This time translation invariance of $\dot\phi_i(t)$ for arbitrary $\delta t$ would explain why the chimera state is observable,
even though the trajectory of the state is neutrally stable.

Based on the numerical results, the chimera state that we observe is not chaotic, in contrast to the finite chimera state
with $\alpha \in (0, \pi/2)$ that is chaotic before collapse to complete 
phase synchronization~\cite{chimera_chaotic, chimera_transient, chimera_spectral, transient_chaotic_exp}.
Recently, several stable chaotic chimera states of finite size have been suggested 
using different types of oscillators~\cite{weak_chimera_chaotic, chimera_persistent1, yscho_prl_2017}.
Along these lines, we may find stable chaotic chimera states of finite size by avoiding the complete phase synchronization of 
non-locally coupled Kuramoto--Sakaguchi oscillators
in arrays for larger $N$~\cite{weak_chimera_chaotic, symmetric_chaotic}.

\section{Basin stability of the chimera state}
\label{sec:basin_chimera}

For $\alpha \in (\pi/2, \pi)$, we measure the fraction of 
random initial conditions $(\phi_1(0),...,\phi_6(0)) \in [0, 2\pi)^6$ that arrives at the chimera state
following Eq.~(\ref{Eq:governing_eq_final}).  
To be specific, we integrate Eq.~(\ref{Eq:governing_eq_final}) up to $t = 10^4$ for each initial condition,
and regard the final state as the chimera state if it satisfies the following two conditions: $\phi_1 = \phi_4$
and $\phi_i \neq \phi_j$ for any pairs $\{i, j\} \in \{1,2,3,5,6\}$ 
(as well as two other equivalent conditions given by the rotational symmetry of the network),
and all $\dot\phi_i$ are periodic functions of $t$.
For the latter, 
we regard each $\dot\phi_i$ as a periodic function if the standard deviation of the distances between two consecutive peak points of the function 
during $9 \times 10^3 \leq t \leq 10^4$ is less than the step-size of $t$ used to integrate Eq.~(\ref{Eq:governing_eq_final}) numerically.
Here, we take $t = 10^4$ for the upper limit of integration to measure basin stability after discarding the initial transients, 
because the chimera state in Fig.~\ref{Fig:chimera_observation}(a) appeared for a time interval of integration shorter than $10^3$
beginning with a random condition.

\begin{figure}[t!]
\includegraphics[width=1.0\linewidth]{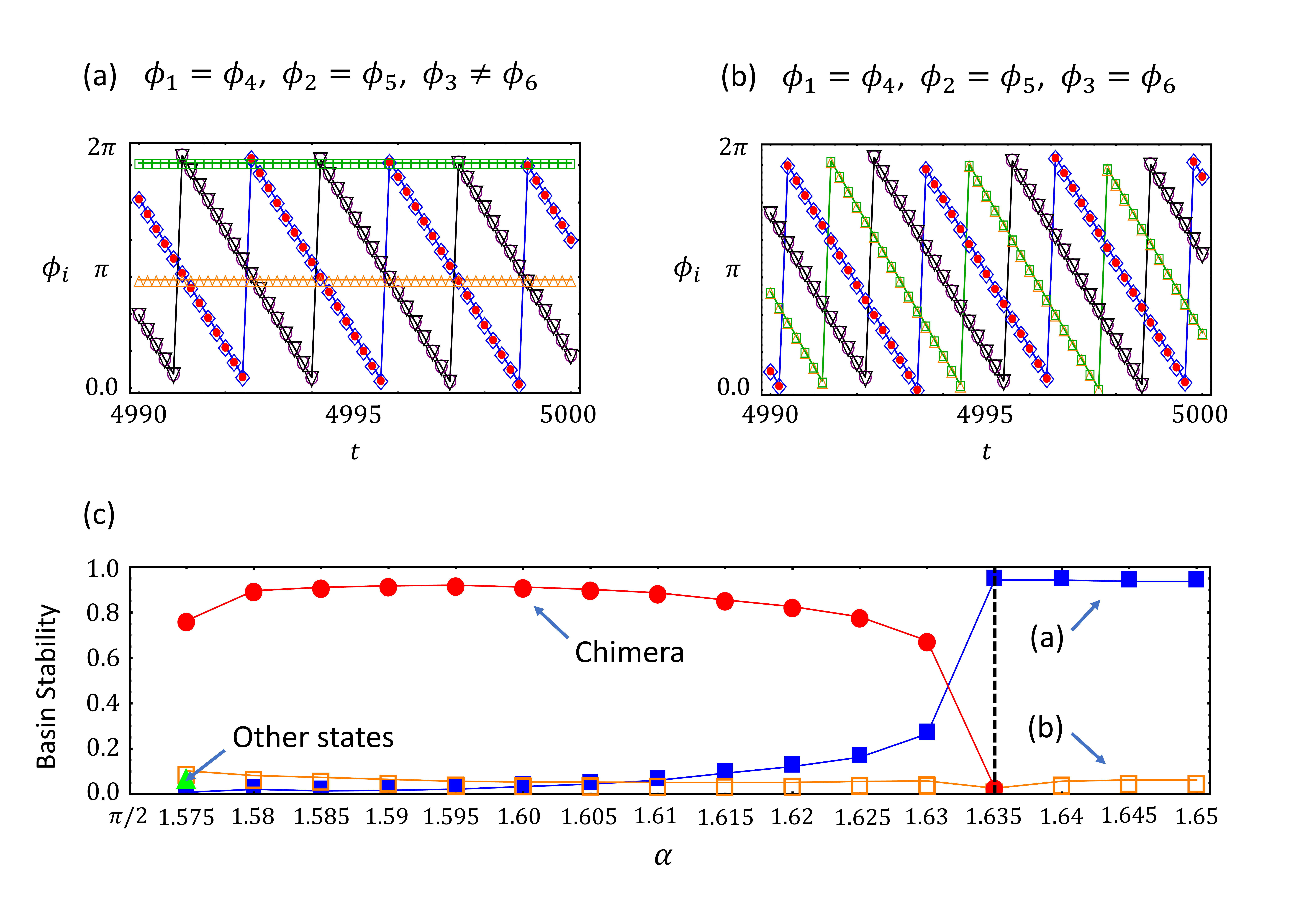}
\caption{Basin stability of the chimera state and other states.
$\phi_1(\diamond)$, $\phi_2(\circ)$, $\phi_3(\oblong)$, $\phi_4(\bullet)$, $\phi_5(\triangledown)$, and $\phi_6(\vartriangle)$ of
(a) a state composed of two synchronous clusters ($\{1, 4\}$, $\{2, 5\}$) and an asynchronous cluster ($\{3, 6\}$), and
(b) a state composed of three synchronous clusters ($\{1,4\}$, $\{2,5\}$, $\{3,6\}$).
(c) Basin stability of the chimera state ($\bullet$) and the two states in (a) $(\blacksquare)$ and (b) $(\square)$ versus $\alpha$.
For each value of $\alpha$, we use $10^4$ random initial conditions.
The vertical dashed line at $1.635$ indicates where the chimera state is no longer observed
in the range of $\alpha$ to the right of the line. 
For each value of $\alpha$, only the symbols of the states with nonzero basin stability are marked.
For $\alpha=1.575$, we observe states other than the chimera state and the two states in (a) and (b) ($\blacktriangle$).}
\label{Fig:chimera_basin}
\end{figure}

We observe the chimera state with a finite probability for $\alpha < 1.64$, 
whereas no chimera state can be observed outside of this range 
as shown in Fig.~\ref{Fig:chimera_basin}(c).
This might be because the basin stability of the chimera state is  exceedingly small or zero outside this range.

In the entire range of $\pi/2 < \alpha < \pi$, we observe two other states
as plotted in Fig.~\ref{Fig:chimera_basin}(a) and (b).
The trajectory of the state in Fig.~\ref{Fig:chimera_basin}(a) is 
($\phi_1=\phi_4 = \omega t + C_1$, $\phi_2=\phi_5 = \omega t + C_1 + \pi$,
$\phi_3 = C_2$, $\phi_6 = C_2 + \pi$),
and that of the state in Fig.~\ref{Fig:chimera_basin}(b)
is given by ($\phi_1=\phi_4=\omega t + C_3$, $\phi_2=\phi_5=\omega t + 2\pi/3 + C_3$, $\phi_3=\phi_6 = \omega t + 4\pi/3 + C_3$)
for arbitrary constants $C_1, C_2, C_3$. 
Here, $\omega = - 2 \textrm {sin}(\alpha)$ is derived for both states.
We obtain the basin stability for these two states
(considering other sets of trajectories 
given by the rotational and reflectional symmetry of the network) as shown in Fig.~\ref{Fig:chimera_basin}(c)
using the same upper limit of integration $t = 10^4$.
We note that these two states are distinct from the chimera state in the sense that they respectively include two and three synchronous clusters,
in contrast to the chimera state having only one synchronous cluster.

\section{Discussion}
\label{sec:discussion}

In this paper, we have discussed phase synchronization $\phi_i = \phi_j$ of oscillators $i \neq j$. 
From the perspective of phase synchronization, we observed a chimera state 
in the network depicted in Fig.~\ref{Fig:chimera_observation}(a), 
where six oscillators are partitioned into a synchronous cluster $C_1=\{1,4\}$ and an asynchronous cluster $C_2=\{2,3,5,6\}$.
Previously, a study~\cite{chaos_weak_chimera1} considered frequency synchronization 
$\Omega_i = \Omega_j$ of oscillators $i \neq j$, where the frequency of each oscillator $i$ is given by
$\Omega_i = \textrm{lim}_{t \rightarrow \infty}\frac{1}{t}\int_{0}^{t}\dot{\phi}_i(t')dt'$.
From the perspective of frequency synchronization, the authors introduced the so-called weak chimera state for oscillators $i,j,k$ in which $\Omega_i \neq \Omega_j$ and $\Omega_i = \Omega_k$.
In the invariant subspace of the three-oscillator quotient system 
$(\phi_1=\phi_4, \phi_2=\phi_6, \phi_3=\phi_5)$ of Eq.~(\ref{Eq:governing_eq_final}) 
with the same network, they reported a weak chimera state 
where $\Omega_2 \neq \Omega_1$ and $\Omega_2 = \Omega_3$.
Such existence of weak chimera states in three-oscillator quotient systems
has recently been understood analytically~\cite{chaos_weak_chimera2}.

In the present work, we numerically measure $\Omega_i=\frac{1}{t} \int^{t}_0 \dot{\phi}_i(t')dt'$ of the chimera state in Fig.~\ref{Fig:chimera_observation}(a)
as $\Omega_i = -1.61081 \pm 0.00001$ $(i = 1,2,4,5)$ and $\Omega_i = -0.05504 \pm 0.00002$ $(i=3,6)$
by integrating $\dot\phi_i$ up to $t = 10^5$.
Based on the obtained values of $\Omega_i$, we assume that the oscillators in the chimera state
might be partitioned into two clusters $\{1,2,4,5\}$ and $\{3,6\}$, where the oscillators in each cluster have the same value of $\Omega_i$.  
Consequently, the chimera state would be a weak chimera state satisfying $\Omega_1 \neq \Omega_3$ and $\Omega_1 = \Omega_2$ 
in the invariant subspace of this five-oscillator quotient system
$(\phi_1=\phi_4, \phi_2, \phi_3, \phi_5, \phi_6)$. We may understand the existence of the chimera state analytically 
by extending the analysis in previous works~\cite{chaos_weak_chimera1, chaos_weak_chimera2}
to the invariant subspace of this five-oscillator quotient system.

Finally, we note that the persistence of the synchronous state of the one subset  
irrespective of the asynchronous phases of the other subset in the chimera state
is related to the invariance of the adjacency matrix (symmetry) 
under permutations within the synchronous subset~\cite{yscho_prl_2017, remote_prl_2013, pecora_ncomm_2014, pecora_sciadv_2016}. 
The number of permutations conserving an adjacency matrix 
usually increases drastically with network size~\cite{symmetry};
therefore, we expect that the formation of synchronous subsets in diverse chimera states in large networks 
can be understood from the perspective of symmetry
under permutations within each subset.

\begin{acknowledgments}
This work was supported by National Research Foundation
of Korea (NRF) Grant No. 2017R1C1B1004292.
\end{acknowledgments}


\end{document}